\documentclass[aps,prl,twocolumn,groupedaddress,showpacs]{revtex4}

\usepackage{graphicx}
\usepackage{colortbl}
\usepackage{epstopdf}

\newcommand{\Rey}{\mathrm{Re}}
\newcommand{\Pe}{\mathrm{Pe}}
\newcommand{\EE}{\mathcal{E}}
\newcommand{\TT}{\mathcal{T}}

\newcommand{\float}[2]{#1\times 10^{#2}}
\begin{document}

\title{Enhanced shear separation for chiral magnetic colloidal aggregates}  
\author{C. I. Mendoza$^{1,2}$, C. M. Marques$^{2}$, F. Thalmann$^{2,\ast}$}
\affiliation{$^{1}$ Instituto de Investigaciones en Materiales,
  Universidad Nacional Aut%
\'{o}noma de M\'{e}xico, Apdo. Postal 70-360, 04510 M\'{e}xico, D.F., Mexico
\\ 
$^{2}$ Institut Charles Sadron, Universit\'e de Strasbourg, CNRS UPR
22, 23 rue du Loess, Strasbourg Cedex, F-67037, France.}
\date{\today}

\begin{abstract}
We study the designing principles of the simplest colloidal propeller,
an architecture built from four identical spheres that can couple
translation with rotation to produce controlled drift motion. By
considering superparamagnetic beads, we show that the simultaneous
action of a magnetic field and a shear flow leads to the migration of
the cluster in the vorticity direction.  We investigate the dependence
of the migration velocity on the geometrical parameters of the
cluster, and find that significant cluster separation can be achieved
under the typical operation conditions of microfluidic devices.
\end{abstract}

\maketitle

%
%
%

The motion and the propulsion of microscopic living organisms, the
development of top-down techniques to carve micrometer-size devices
like pumps, switches and motors, the supramolecular chemistry quest
for molecular motors and the new bottom-up techniques that emerge from
the Soft Nanosciences, have lead to a reborn interest for the
hydrodynamics at low Reynolds numbers $\Rey$.  
In the microscopic world of fluids where typical dimensions $\ell$ are
of the order of, or smaller than a few micrometers, $\ell\lesssim
10\ \mu$m, where the velocities $v$ are smaller than a few micrometers
per second, $v\lesssim 10\ \mu$m.s$^{-1}$, the ratio $\Rey= \ell v
\rho/ \eta$ between inertial and viscous forces is extremely small,
$\Rey\lesssim 10^{-4}$. 
At these scales, inertia is irrelevant: for all practical purposes
when the applied forces stop, the movement stops. Designing feasible
and efficient propulsion devices for low Reynolds numbers is arguably
one of the main current scientific challenges in the Nanosciences.
Inspired by recent progress in the self-assembling small numbers of
colloidal particles into aggregates with well defined
geometries~\cite{2003_manoharan_pine}, and motivated by the
possibility of manipulating colloidal aggregates of superparamagnetic
beads with external magnetic fields~\cite{2005_dreyfus_bibette}, we
propose and study in this Communication the simplest feasible colloidal
aggregate that will behave as a small propeller, capable of
transducing rotation into translation and vice-versa.

The propeller effect consists in rotating at constant angular velocity
$\omega$ a chiral aggregate around a fixed orientation axis,
conventionally named $z$, and subsequently referred as the
  vertical axis, causing a net constant drift velocity $v_z$
along $z$. The propeller efficiency of an aggregate of size $L$ can be
expressed as a dimensionless \textit{pitch} $P_{p}=2\pi v_z/(L\omega)$
that, in the Stokes flow approximation, provides a size independent
characterization of its ability to couple rotational and translational
motions. Spatial preorientation of the aggregate is required to
observe a sizeable linear propeller effect, otherwise strongly
suppressed by the free rotational Brownian motion. We show below that
this can be achieved with a constant uniform magnetic field acting on
superparamagnetic colloids.

Rotation of the aggregate can, in principle, be enforced by means of
circular polarized
light~\cite{2004_Bishop_RubinszteinDunlop}. Alternatively, a
precessing magnetic field with small but finite tilt angle $\beta$
could also be used, but only with low efficiency
(\textit{i.e.}$\sim\beta^2$)~\cite{MTM_in_preparation}.  In any case,
it is simpler to rely on flows to rotate the clusters, leading
naturally to a shear induced vertical drift in Couette geometry.

It was recognized some time ago that the migration under shear flow of
a colloidal cluster with randomized orientation only shows a cubic
dependence in the shear rate $\dot{\gamma}$~\cite{2005_doi_makino_1}.
Shear separation becomes linear in $\dot{\gamma}$ as soon as
aggregates get preoriented with respect to the shear flow, with $z$
parallel to the vorticity direction. We introduce the shear separation
efficiency ratio $P_s=-4\pi v_z/(L\dot\gamma)$ as a counterpart of
$P_{p}$, based on a correspondence $\omega=-\dot{\gamma}/2$ between
angular velocity and shear rate. Propeller and shear separation are
related, but were found inequivalent in the cases investigated so far,
and comparing $P_p$ and $P_s$ is an important issue. Strong constant
shear rates can be experimentally applied, \textit{e.g.}  using soft
lithography microfluidic
devices~\cite{2009_Marcos_Stocker,2010_Eichhorn} and shear separation
has the potential to become a method of choice for partitioning
aggregates according to their chirality.

We first discuss the preorientation of the colloid aggregates by the
applied field. Each superparamagnetic bead $i$ making up the aggregate
acquires an induced magnetic moment $\vec{m}_i$ and the equilibrium
position of the aggregate is the one minimizing the sum $-\sum_i
\vec{m}_i \cdot\vec{B}/2$, where $\vec{B}$ stands for the uniform,
gradientless, magnetic induction in the absence of colloids. We also
assume that the magnetic content of the beads is homogeneous, and that
the aggregate is rigid. The actual determination of the best
orientation of the aggregate depends on the strength of mutual
magnetic moments interaction, controlled by a dimensionless parameter
$\varepsilon=\mu_0\alpha_b/(32\pi a^3)$, with $\mu_0$ the
permeability, $\alpha_b$ the magnetic polarizability, \textit{i.e} the
ratio between magnetization $m$ and induction $B$, and $a$ the radius
of the beads. Provided that \textit{i}$)$ the applied magnetic field
$||\vec{B}||$ is less than a critical value $B_c$ and does not bring
$\vec{m}_i$ close to its saturation value, \textit{ii}$)$ the time
variations of $\vec{B}$ are slower than the N\'eel relaxation time
$\tau_N$ of the magnetic particles and \textit{iii}$)$ the interaction
parameter $\varepsilon$ remains smaller than unity, then the magnetic
energy reduces to $E_m(\vec{B},\Omega)=\sum_{i< j}^n V_{ij}$ with a
dipole-dipole interaction equal to $V_{ij}= 8\alpha_b
a^3\varepsilon\vec{B}^2 r_{ij}^{-3}(1-3\cos^2\theta_{ij})$,
$\theta_{ij}$ the angle between the magnetic field and the vector
$\vec r_{ij}$ connecting beads $i$ and $j$, and $\Omega$ a set of
Euler angles fixing the orientation of the cluster. This
  dipolar energy is the leading term of an expansion in powers of
  $\varepsilon$ that fully takes into account the magnetic
  interactions between beads.

The quadratic dependence of $E_m(\vec{B},\Omega)$ in $\vec{B}$ at
fixed $\Omega$ leads to a unique optimal orientation of the aggregate
with respect to $\vec B$ if the spectrum of the quadratic form has a
single lowest eigenvalue, which is the case for generic, non-symmetric
aggregates. Otherwise there may be a degenerate subspace of low energy
configurations, while aggregates with high symmetry end up being
totally insensitive to the applied field, as it happens \textit{e.g.}
for a regular four beads tetrahedron.

For dimers, the preferred orientation is along the field direction;
for trimers, as for any planar aggregate, the plane of symmetry of the
cluster contains the field direction. These are particular instances
of more general orientation rules which are strongly constrained by
the symmetry properties of the aggregate. Assuming that a given
aggregate possesses a well defined orientation, then any symmetry
plane $\sigma$ of the aggregate is either normal to $\vec{B}$ or
contains it, any rotation axis $C_n$ of order $n\geq 3$ is parallel to
$\vec{B}$, and any rotation axis $C_2$ ends up being parallel or
orthogonal to $\vec{B}$. We assume from now on that our colloidal
aggregates possess a well defined orientation axis and rotate around
it. In the case of shear induced migration, the vorticity of the shear
flow should be aligned parallel with $\vec{B}$ in order to facilitate
the rotation of the aggregate. Flows with vorticity perpendicular to
$\vec{B}$ induce torques working against the magnetic stability of the
aggregate, ending up with no or little rotation, and are not
considered further in the present approach.

Let us consider the linear relation $v_z=\kappa \omega$ between drift
and angular velocities of an aggregate with constrained rotation axis
$z$. If an inversion center of the aggregate exists, this inversion
symmetry preserve both $\kappa$ and $\omega$ while reversing $v_z$. If
a mirror symmetry plane containing the axis $z$ exists, it preserves
$\kappa$ and $v_z$, but reverses $\omega$. If the mirror plane is
orthogonal to the $z$ direction, it preserves $\kappa$ and $\omega$,
and reverses $v_z$. Therefore, $\kappa$ vanishes for all non-chiral
aggregates. The same consideration holds when a torque $\TT$ is
applied instead of a constant angular velocity $\omega$. Dimers and
trimers are \textit{de facto} ruled out as they possess a mirror plane
containing the field. It requires at least four beads to build a
propelling aggregate, the only point-symmetry groups allowed for
propelling aggregates are, according to Schoenflies terminology,
C$_1$, C$_n$ and D$_n$ ($n\geq 2$)~\cite{book_Carter}.

\begin{figure}
\includegraphics[width=2.9in]{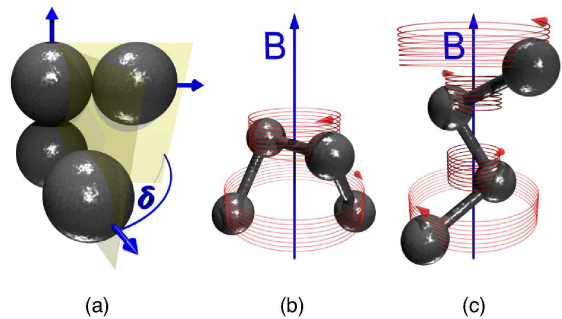}
\caption{%
Sketch of the colloidal propeller, made of identical superparamagnetic
beads. For displaying purposes only, the beads are represented by
spheres of a diameter smaller that the arm length in (b) and (c). The
propeller consists of a main axis and two perpendicular arms that form
an angle $\delta$ (a). For small values of $\delta$, one stable
equilibrium position under vertical magnetic field, that we name
$\Lambda$ orientation, is shown in (b), the other being obtained by a
rotation of 180$^\circ$ around the main axis. For larger values of
$\delta$ the equilibrium position that we name $\Sigma$ orientation is
shown in (c).%
}%
\label{fig:aggregates}
\end{figure}

We will now show that some tetrameric assemblies display indeed
propeller and shear separation properties, and compute the magnitude
of these effects.  We consider an aggregate of four identical beads
rigidly connected, as depicted in Fig.~(\ref{fig:aggregates}.a). In
this configuration, bead one and bead two at a diameter distance
$L=2a$ define the main axis. The third bead, connected to bead one
defines a lower arm of length $L$ perpendicular to the main axis, and
the fourth bead, connected to bead number two defines an upper arm of
length $L$, also perpendicular to the main axis. The upper and lower
arms are separated by an angle $\delta$.  Aggregates with angles
$\delta$ and $-\delta$ only differ in their chirality, also called
enantiomers. We consider $L$ as the characteristic size of the
aggregate.  

The equilibrium position of such cluster under magnetic field depends
on $\delta$. For small values $\delta< \delta_c$, the typical
orientation of the cluster that we name orientation $\Lambda$ is
displayed in Fig.~(\ref{fig:aggregates}.b). One observes that the
field direction is along the (unique) axis of symmetry $C_2$ of the
tetramer. For larger values of $\delta$, the stable configuration,
that we call orientation $\Sigma$, adopts an helical conformation as
shown in Fig.~(\ref{fig:aggregates}.c), with $\vec{B}$ perpendicular
to the $C_2$ axis, in accordance with the above orientation rules. The
$\Lambda$-$\Sigma$ stability exchange, caused by an eigenvalues
crossing of $E_m(\vec{B},\Omega)$ takes place around $\delta_c
\simeq 66.4^{\circ}$.

Within the Rotne-Prager-Yamakawa dipolar
approximation~\cite{1969_Rotne_Prager} the velocity of the particle
located at $\mathbf{R}_{m}$ is given by the hydrodynamic relation
$\mathbf{V}_{m}=\sum_{n}\mathbf{H}_{mn}\cdot \mathbf{F}_{n}$, where
$\mathbf{F}_{n}$ are the forces acting on particles and the
mobility matrix $\mathbf{H}_{mn}$ is approximated by
$\mathbf{H}_{nn}=\mathbf{I}/(6\pi\eta a)$ and
$\mathbf{H}_{mn}=\mathbf{H}(\mathbf{R}_{m}-\mathbf{R}_{n}) \ {\rm
  for\ }m\neq n$ with $\mathbf{H}\left( \mathbf{r}\right) =%
1/(8\pi\eta r^3)\left(r^2\mathbf{1}+\mathbf{r}\mathbf{r}+%
2a^2(\mathbf{1}/3-\mathbf{r}\mathbf{r}/r^2)\right)$. The validity of
such modeling of cluster hydrodynamics was investigated
in~\cite{2007_GarciadelaTorre_Ortega} and suggests an accuracy within
30\% for the rotational drag coefficient. The rigid cluster movement
is reducible to three translational and three rotational degrees of
freedom collectively denoted as $\mathcal{Q}_{\alpha}$,
$\alpha=x,y,z,\phi,\theta,\psi$, to which are associated generalized
forces and torques $\mathcal{F}_{\alpha}$. Internal constraints can be
eliminated by suitably projecting the inverse hydrodynamic tensor
$\mathbf{H}^{-1}$ over the collective displacements
$\partial\mathbf{R}_m/\partial{Q}_{\alpha}$, leading
to~\cite{2001_doi_edwards}:
\begin{equation} 
\dot{Q}_{\alpha}=\sum_{\gamma=1,\ldots 6}\mathbf{h}^{(6)}_{\alpha,\gamma}
\mathcal{F}_{\gamma}+\mathcal{U}_{\alpha} \label{eq:formalism-general}  
\end{equation}
where $\mathbf{h}^{(6)}_{\alpha,\gamma}$ is a $6\times 6$ symmetric
mobility matrix restricted to the collective coordinates obeying
$\left( \mathbf{h}^{-1}\right) _{ab}=\sum_{m,n}\partial \mathbf{R}_{n}/\partial
Q_{a}\cdot \left( \mathbf{H}^{-1}\right) _{nm}\cdot \partial
\mathbf{R}_{m}/\partial Q_{b}$, $\mathcal{F}_{\gamma}$ are the
external forces and torques components, and $\mathcal{U}_{\alpha}$ the
generalized components of the streaming flow. In the ideal propeller
case, the motion reduces to only two coordinates $\theta$ and $z$,
standing respectively for the rotation around $z$ axis, and
translation parallel to it. In the absence of flow, one gets:
\begin{equation}
\dot{z}=\mathbf{h}^{(2)}_{z,z} F_z + \mathbf{h}^{(2)}_{z,\theta}\TT_{\theta};\; 
\dot{\theta}=\mathbf{h}^{(2)}_{\theta,z} F_z + \mathbf{h}^{(2)}_{\theta,\theta}
\TT_{\theta}
\label{eq:formalism-propeller}
\end{equation}
with $F_z$ the force, $\TT_{\theta}$ the torque and $\mathbf{h}^{(2)}$
a $2\times 2$ symmetric matrix. Eq.~(\ref{eq:formalism-propeller})
suffices to discuss the properties of propeller configurations, while
eq.~(\ref{eq:formalism-general}) must be numerically integrated when
considering shear separation experiments or rotation under precessing
magnetic field. Numerical integration of these equations by a
fourth-order Runge-Kutta method \cite{1985_arfken} provides complete
solutions for the time evolution of the center of mass and orientation
of the propeller.

\begin{figure}
\includegraphics[width=2.90in]{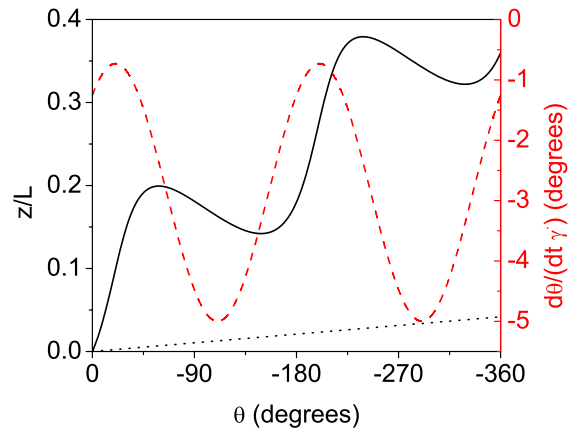}
\caption{(Color online) Angular velocity (red dashed line) and
  vertical position of the center of mass (black full line) of the
  propeller as a function of time while it turns in a shear flow. The
  straight dotted line represent the vertical motion of a propeller
  rotating with same angular velocity, $\delta=40^{\circ}$.}
\label{fig:movementcm}
\end{figure}

Applying a torque $\TT_{\theta}$ and no force $F_z$ results in an
helical trajectory $\theta(t)$, $z(t)$ associated with a pitch
$P_{p}=(2\pi/L) \dot{z}/\dot{\theta}= (2\pi/L)
\mathbf{h}^{(2)}_{\theta,z}/\mathbf{h}^{(2)}_{\theta,\theta}$ which
confirms the geometrical, time-reparametrization invariant of the
motion. In a Couette flow with vorticity parallel to the magnetic
field, the aggregate rotates producing an oscillatory vertical
movement with a net macroscopic migration along the vorticity
direction. The angular velocity of the propeller also oscillates in
time, as shown in Fig.~(\ref{fig:movementcm}), depending on the
relative attitude of the cluster with respect to the anisotropic shear
flow. Curiously enough, the faster vertical motion coincide with the
slowest angular velocity part of the trajectory. This can be
understood if one observe that an anisotropic shear flow can exert
a lift force on a non rotating aggregate. 

\begin{figure}[tbp]
\includegraphics[width=2.90in]{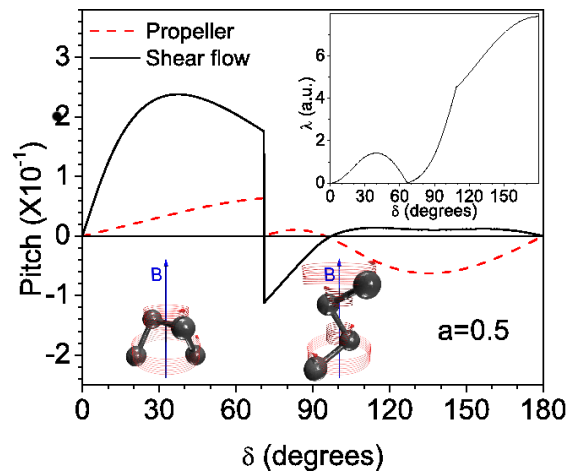}
\caption{(Color online) Pitches $P_s$ (shear flow, full black line)
  and $P_{p}$ (propeller, dashed red curve) as a function of the
  angle $\delta$ for touching beads $L=2a$. $\Lambda$ and $\Sigma$
  orientations are depicted on the graph, separated by a critical
  value $\delta_c$. Inset: stability $\lambda_g$ of the magnetically
  oriented aggregate.}
\label{fig:pitch}
\end{figure}

We show $P_p$ as a function of $\delta$ in Fig.~(\ref{fig:pitch}).
For the two points with no chirality ($\delta =0^\circ$ and
$180^\circ$) the pitch vanishes as expected.  $P_{p}$ is also seen
numerically to vanish around $\delta_d \simeq 95^{\circ}$, thus
separating the $\Sigma$ regime in two regions with pitches of opposite
sign. A maximum of $P_{p}\simeq 0.05 $ corresponds to $\delta_c\simeq
66^{\circ}$ and a minimum of $P_{p}\simeq -0.05$ obtained for
$\delta_m\simeq 130^{\circ}$. As a complete rotation of the cluster
corresponds to a vertical shift of $\Delta z=P_{p} L$, it requires
about 20 rotations of our cluster to move upwards over its own
size. 


The shear flow pitch $P_s$ correlates well with $P_{p}$ in the
$\Lambda$ orientation, but has a sign opposite to $P_{p}$ in the
$\Sigma$ orientation. In particular, $P_s$ and $P_{p}$ apparently
vanish for the same value $\delta_d$, in agreement with a general
argument stating that $P_{p}=0 \Rightarrow P_s=0$, without
reciprocity~\cite{MTM_in_preparation}.  The propeller efficiency in
the absence of external flow is overall lesser than the shear slow
efficiency in $\Lambda$ orientation while it is stronger in $\Sigma$
orientation. The relative sign of the two effects depends on the
details of the colloidal cluster.
 
The inset of Fig.~\ref{fig:pitch} shows the stability
$\lambda_{g}(\delta)$ of the magnetic field induced orientation of the
cluster, as a function of $\delta$. $\lambda_g$ is defined as twice
the difference between the two smallest eigenvalues of the quadratic
form $E_m(\vec{B},\Omega)$, with $8\alpha_b\varepsilon=1$. As a
result, the depth of the magnetic induced energy minimum and the
maximal equilibrium restoring magnetic torque can both be expressed as
$\EE_{mag}=\TT_{mag}=4\varepsilon\alpha_b\lambda_{g}\vec{B}^2$.

We have so far neglected thermal forces in our analysis. The
competition between drift and diffusion defines a Peclet
number $\Pe = |v_z| L/ D_t$, with a diffusion coefficient
$D_t=k_B T/(6\pi\eta L)$. In the sheared situation
$v_z=-P_s\dot{\gamma}L/4\pi$ and the ratio between vertical and
  Brownian displacements scale as
  $(\dot{\gamma}\mathrm{Pe}P_s\,t/8\pi)^{1/2}$, defining a
  characteristic time $t_c=8\pi/(\dot{\gamma}\mathrm{Pe}P_s)$ beyond
  which the drift displacement dominates.

Let us illustrate with numbers these phenomena in the case of a shear
experiment. Many common superparamagnetic beads contain crystallites
of $\gamma\mathrm{Fe}_2\mathrm{O}_3$ with estimated radius $r_c\simeq
10$~nm~\cite{2007_Jeong}. Each one of these magnetic grains bears a
thermally fluctuating moment of magnitude $m_{p} =
\float{1.9}{-18}~\mathrm{A.m}^2$ with random orientation.  This sets
the saturation field to $B_c=k_BT/m_{p}\sim 2.2$~mT and the linear
response polarizability of crystallites amount to $\alpha_p \sim
\float{3}{-16}\mathrm{A.m}^2.\mathrm{T}^{-1}$. The N\'eel relaxation
time was experimentally found close to $\tau_N\sim
\float{4}{-5}$s~\cite{1997_Fannin_Bibette}.  The polarizability
$\alpha_b$ and coupling parameter $\varepsilon$ of a bead both depend
on the magnetic material volume fraction $\Phi$, leading respectively
to $\alpha_b = \alpha_p (a^3/r_c^3)\Phi$ and $\varepsilon\simeq 3.6\,
\Phi$.

The dipolar energy used in our approach is correct for the moderately
magnetic beads ($\Phi=0.02,\, \varepsilon \simeq 0.07$) used
in~\cite{1997_Zahn_Maret} while corrections to the magnetic energy are
required for the stronger magnetic beads ($\Phi\simeq 0.6,
\varepsilon\simeq 2.2$) studied in~\cite{1997_Fannin_Bibette}.  The
magnetic orientation energy must excede the thermal energy $k_B T$.
Our calculation is restricted to the linear polarization regime,
leading to a magnetic stabilizing energy of order
$\EE_{mag}\sim\float{4.8}{6}a^3 \Phi^2\lambda_{g}k_BT$, with $a$
expressed in micrometers. The magnetic torque $\TT_{mag}$ must match
the viscous torque $\TT_{drag}\sim 4\pi \eta L^3 \dot{\gamma}$ exerted
by the flow, with a dynamic viscosity $\eta=10^{-3}$Pa.s in the case
of dilute water solutions. This, at room temperature, limits the shear
rate to $\dot{\gamma}_{max} \sim \float{2.0}{5}\Phi^2\,\lambda_{g}$
irrespective of cluster size $L$. The larger drift velocity $v_z$
consistent with these constraints scales as $V_{max}=P_s\, L\,
\dot{\gamma}_{max}/(4\pi)$, or equivalently $V_{max} =
\float{3.1}{4}P_s a\Phi^2 \lambda_{g}$, with a corresponding Pe
scaling as $\Pe = \float{5.6}{5} P_s a^3\Phi^2\lambda_g$ ($a$ in
micrometers).

For micron sized beads with moderate content in magnetic material
($\Phi=0.1$), assembled with an angle $\delta=40^{\circ}$, one reads
from Fig.~\ref{fig:pitch} the values $\lambda_g\sim 1.4$ and
$P_s=0.25$. One obtains a maximal shear rate $\dot{\gamma}_{max}\sim
2800~\mathrm{s}^{-1}$, a maximal drift velocity $V_{max}=
110~\mu$m.s$^{-1}$, $t_c=18~\mu\mathrm{s}$ and Pe=2000. This should be
fast enough to induce some enantiomer separation using a microfluidic
chip like the one described in~\cite{2009_Marcos_Stocker} with a
magnetic field aligned parallel to the chip and orthogonal to the main
direction of the flow. The scaling $\mathrm{Pe}\sim a^{-3}$ limits our
results to sizes $a= 80~\mathrm{nm}$ below which $\mathrm{Pe}\leq
1$ and Brownian forces cannot be neglected any longer.

In summary, we have applied an hydrodynamic interaction formalism for
studying the rotational-translational coupling for a chiral propeller
made of four identical beads. This coupling can be optimized by
changing the geometry of the aggregate. Chiral superparamagnetic
colloidal enantiomers can be oriented and separated under the
simultaneous action of a magnetic field and a shear flow even at
relatively low shear rates since they migrate in opposite directions
with velocity proportional to the applied shear.  Propeller and shear
separation properties are distinct features, their relative sign may
change depending on the details of the cluster, and are not simply
related to the handedness of the chiral
aggregate~\cite{1999_Harris_Lubensky}. Quantitative estimates of the
magnitude of this magnetic field assisted drift are given that support
the feasibility of the experiment with microfluidic devices.

CIM acknowledges the Chemistry Department of the Centre National de la
Recherche Scientifique for a visiting grant and partial financial support by
DGAPA-PAPIIT Grant No. IN-115010. Enlightening discussions are
acknowledged with Paul Chaikin and Thomas Ebbesen.

$^{\ast }$fabrice.thalmann@ics-cnrs.unistra.fr

\bibliographystyle{plain}

\clearpage

\end{document}